\documentclass[11pt]{article}

\usepackage{graphicx}
\usepackage[margin=0.5in]{geometry}
\usepackage{bm}
\usepackage{amsthm}
\usepackage{amscd}
\usepackage{amsbsy}
\usepackage{appendix}
\usepackage{amsmath}
\usepackage{amsfonts}
\usepackage{amssymb}
\usepackage{calligra}
\usepackage{xcolor}
\usepackage{float}
\usepackage[T1]{fontenc}
\usepackage{multirow}
\usepackage{mathrsfs}
\usepackage{mathtools}
\usepackage{epstopdf}
\usepackage{psfrag}
\usepackage{subfigure}
\usepackage{booktabs}
\usepackage{listings}
\usepackage{longtable}
\usepackage{latexsym}
\usepackage{arydshln}
\usepackage{enumerate}
\usepackage{epsfig}
\usepackage{cite}
\usepackage{verbatim}
\usepackage{overpic}
\usepackage{abstract}
\usepackage{extarrows}
\usepackage[pagewise]{lineno}

\usepackage[noend]{algpseudocode}
\usepackage{algorithmicx,algorithm}

\allowdisplaybreaks[4]

\date{}
\title{Algorithmic Averaging for Studying Periodic Orbits of Planar Differential Systems}
\author{Bo Huang$^{\text{a,b}}$\\
	\it\footnotesize $^{\text{a}}$LMIB -- School of Mathematical Sciences,
Beihang University, Beijing 100191, China \\
	\it\footnotesize $^{\text{b}}$Courant Institute of Mathematical Sciences, New York University, New York 10012, USA\\
	\it\footnotesize bohuang0407@buaa.edu.cn}

\newtheorem {theorem*}{Theorem}
\newtheorem{theorem} {Theorem}
\newtheorem{example}{Example}
\newtheorem{definition}{Definition}

\newtheorem{proposition}{Proposition}
\newtheorem{lemma}{Lemma}
\newtheorem{corollary}{Corollary}
\newtheorem{remark}{Remark}

\newtheorem{open problem} {Open problem}

\numberwithin{equation}{section}

\usepackage[colorlinks,
            CJKbookmarks=true,
            linkcolor=blue,
            anchorcolor=red,
             urlcolor=blue,
            citecolor=blue
            ]{hyperref}

\begin{document}
\maketitle
\noindent {\bf Abstract.}  One of the main open problems in the qualitative theory of real planar differential systems is the study of limit cycles. In this article, we present an algorithmic approach for detecting how many limit cycles can bifurcate from the periodic orbits of a given polynomial differential center when it is perturbed inside a class of polynomial differential systems via the averaging method. We propose four symbolic algorithms to implement the averaging method. The first algorithm is based on the change of polar coordinates that allows one to transform a considered differential system to the normal form of averaging. The second algorithm is used to derive the solutions of certain differential systems associated to the unperturbed term of the normal of averaging. The third algorithm exploits the partial Bell polynomials and allows one to compute the integral formula of the averaged functions at any order. The last algorithm is based on the aforementioned algorithms and determines the exact expressions of the averaged functions for the considered differential systems. The implementation of our algorithms is discussed and evaluated using several examples. The experimental results have extended the existing relevant results for certain classes of differential systems.


\smallskip

\noindent {\bf Math Subject Classification (2010).} 34C07; 37G15; 68W30.

\smallskip


\section{Introduction}

We deal with polynomial differential systems in $\mathbb{R}^2$ of the form
\begin{equation}\label{eq2.0}
\begin{split}
\frac{dx}{dt}=\dot{x}=f_n(x,y),\quad\frac{dy}{dt}=\dot{y}=g_n(x,y),
\end{split}
\end{equation}
where $n$ is the maximum degree of the polynomials $f$ and $g$. As we knew, the second part of the 16th Hilbert's problem \cite{DH00,YI02} asks for ``the maximal number $H(n)$ and relative configurations of limit cycles'' for the differential system \eqref{eq2.0}. Here $H(n)$ is called the Hilbert number. The problem is still open even for $n=2$. However, there have been many interesting results on the lower bound of $H(n)$ for $n\geq2$: it is shown in \cite{lm79,s80} that $H(2)\geq4$ and $H(3)\geq13$ in \cite{ccj09}. In \cite{cn95}, it is proved that $H(n)$ grows at least as rapidly as $n^2\log n$. For the latest development about $H(n)$, we refer the reader to \cite{cl07,jl03,hl12}.


We recall that a limit cycle of the differential system \eqref{eq2.0} is an isolated periodic orbit of the system. One of the best ways of producing limit cycles is by perturbing a differential system which has a center. In this case the perturbed system displays limit cycles that bifurcate, either from the center (having the so-called Hopf bifurcation), or from some of the periodic orbits surrounding the center, see the book of Christopher-Li \cite{cl07} and the references cited therein.

Usually, a limit cycle which bifurcates from a center equilibrium point is called a \textit{small amplitude limit cycle}, and a \textit{medium amplitude limit cycle} is one which bifurcates from a periodic orbit surrounding a center (see \cite{jacm10,jj15}). Note that the notation of ``large'' limit cycle may occur in several situations in the literature, see \cite{mth04,ym12}. In the past seven decades, many researchers have considered the small amplitude limit cycles and obtained many results (e.g., \cite{isk44,nn54,ngl88,dw90,pyah05}). Over the years, a number of algebraic methods and algorithms have been developed (e.g., \cite{ngs88,dw91,vgr93,gt01,yc08,mjp09}) based on the tools of Liapunov constants or Melnikov function.

In our recent work \cite{hy19}, we provide an algorithmic approach to small amplitude limit cycles of nonlinear differential systems by the averaging method, and give an upper bound of the number of zeros of the averaged functions for the general class of perturbed differential systems (\cite{hy19}, Thm. 3.1). The goal of this paper is to extend our algorithmic approach to study the maximal number of medium amplitude limit cycles that bifurcate from some periodic orbits surrounding the centers of the unperturbed systems. The main technique is based on the general form of the averaging method for planar differential systems.

The method of averaging is an important tool to study the existence of isolated periodic solutions of nonlinear differential systems in the presence of a small parameter. It can be used to find a lower bound of the Hilbert number $H(n)$ for certain differential systems. The method has a long history that started with the classical works of Lagrange and Laplace, who provided an intuitive justification of the method. The first formalization of this theory was done in 1928 by Fatou. Important practical and theoretical contributions to the averaging method were made in the 1930's by Bogoliubov-Krylov, and in 1945 by Bogoliubov. The ideas of averaging method have extended in several directions for finite and infinite dimensional differentiable systems. For a modern exposition of this subject, see the books of Sanders-Verhulst-Murdock \cite{svm07} and Llibre-Moeckel-Sim\'o \cite{jrc15}.

The averaging method provides a straightforward calculation approach to determine the number of limit cycles that bifurcate from some periodic orbits of the regarded particular class of differential systems. However, in practice, the evaluation of the averaged functions is a computational problem that requires powerful computerized resources. Moreover, the computational complexity grows very fast with the averaging order. Our objective in this paper is to present an algorithmic approach to develop the averaging method at any order and to further study the number of medium amplitude limit cycles for nonlinear differential systems.

In general, to obtain analytically periodic solutions of a differential system is a very difficult problem, many times a problem impossible to solve. As we shall see when we can apply the averaging method, this difficult problem is reduced to finding the zeros of a nonlinear function in an open interval of $\mathbb{R}$, i.e., now the problem has the same difficulty as the problem of finding the singular or equilibrium points of a differential system.

The structure of our paper is as follows. In Section \ref{sect2}, we introduce the basic results on the averaging method for planar differential systems. We give our algorithms and briefly describe their implementation in Maple in Section \ref{sect3}. Its application is illustrated in Section \ref{sect4} using several examples including a class of generalized Kukles polynomial differential systems and certain differential systems with uniform isochronous centers of degrees 3 and 4. Finally, a conclusion is provided in Section \ref{sect5}. The Maple code of the algorithms can be download from {\textcolor{blue}{\underline{\url{https://github.com/Bo-Math/limit-cycle}}}}.


In view of space limitation, we present the explicit formulae of the $k$-th order averaged function up to $k=5$ in Appendix \ref{A}. The proof of Corollary \ref{cork.1} is moved to Appendix \ref{A-B}. Some experimental results on certain differential systems with uniform isochronous centers are found in Appendix \ref{B}. 

\section{Main Results of the Averaging Method} \label{sect2}
In this section we introduce the basic theory of the averaging method. We consider the following polynomial differential system of degree $n_1$
\begin{equation}\label{eq2.02}
\begin{split}
\dot{x}=P(x,y),\quad\dot{y}=Q(x,y)
\end{split}
\end{equation}
having a center at the point $\bar{x}\in\mathbb{R}^2$. Without loss of generality we can assume that the center $\bar{x}$ of system \eqref{eq2.02} is the origin of coordinates. The following definition is due to Poincar\'e (\cite{jcm99}, Sect. 2).
\begin{definition}
We say that an isolated singular point $\bar{x}$ of \eqref{eq2.02} is a \textit{center} if there exists a neighbourhood of $\bar{x}$, such that every orbit in this neighbourhood is a cycle surrounding $\bar{x}$.
\end{definition}

\begin{remark}
Determining the conditions on the parameters under which the origin for system \eqref{eq2.02} is a center is the well-known \textit{center problem}, see \cite{vd09,acj17}. There are many partial results for the centers of system \eqref{eq2.02} of degree $n_1\geq2$. Unfortunately, at present, we are very far from obtaining the classification of all the centers of cubic polynomial differential systems. In general, the huge number of computations necessary for obtaining complete classification becomes the central problem which is computationally intractable, see for instances, \cite{jgxs04} and the references cited therein.
\end{remark}


Now consider the perturbations of \eqref{eq2.02} of the form
\begin{equation}\label{eq2.04}
\begin{split}
\dot{x}&=P(x,y)+p(x,y,\varepsilon),\\
\dot{y}&=Q(x,y)+q(x,y,\varepsilon),
\end{split}
\end{equation}
where the polynomials $p$, $q$ are of degree at most $n_2$ (usually $n_2\geq n_1$) in $x$ and $y$, and $\varepsilon$ is a small parameter. We are interested in the maximum number of medium amplitude limit cycles of \eqref{eq2.04} for $|\varepsilon|>0$ sufficiently small, which bifurcate from some periodic orbits surrounding the centers of system \eqref{eq2.02}.

Usually, the averaging method deals with planar differential systems in the following normal form
\begin{equation}\label{eq2.01}
\begin{split}
\frac{dr}{d\theta}=\sum_{i=0}^k\varepsilon^iF_i(\theta,r)+\varepsilon^{k+1}R(\theta,r,\varepsilon),
\end{split}
\end{equation}
where $F_i: \mathbb{R}\times D\rightarrow\mathbb{R}$ for $i=0,1,\ldots,k$, and $R:\mathbb{R}\times D\times(-\varepsilon_0,\varepsilon_0)\rightarrow\mathbb{R}$ are $\mathcal{C}^{k}$ functions, $2\pi$-periodic in the first variable, being $D$ an open and bounded interval of $(0,\infty)$, and $\varepsilon_0$ is a small parameter. As one of the main hypotheses, it is assumed that $r(\theta,z)$ is a $2\pi$-periodic solution of the unperturbed differential system $dr/d\theta=F_0(\theta,r)$, for every initial condition $r(0,z)=z\in D$.

The averaging method consists in defining a collection of functions $f_i: D\rightarrow\mathbb{R}$, called the $i$-th order averaged functions, for $i=1,2,\ldots,k$, which control (their simple zeros control), for $\varepsilon$ sufficiently small, the isolated periodic solutions of the differential system \eqref{eq2.01}. In Llibre-Novaes-Teixeira \cite{jdm14} it has been established that
\begin{equation}\label{eq2.01.2}
\begin{split}
f_i(z)=\frac{y_i(2\pi,z)}{i!},
\end{split}
\end{equation}
where $y_i: \mathbb{R}\times D\rightarrow\mathbb{R}$, for $i=1,2,\ldots,k$, is defined recursively by the following integral equations
\begin{equation}\label{eq2.01.3}
\begin{split}
y_1(\theta,z)&=\int_0^{\theta}\Big(F_1(s,r(s,z))+\partial F_0(s,r(s,z))y_1(s,z)\Big)ds,\\
y_i(\theta,z)&=i!\int_0^{\theta}\Bigg(F_i(s,r(s,z))+\sum_{\ell=1}^i\sum_{S_{\ell}}\frac{1}{b_1!b_2!2!^{b_2}\cdots b_{\ell}!\ell!^{b_{\ell}}}\\
&\quad\cdot\partial^LF_{i-\ell}(s,r(s,z))\prod_{j=1}^{\ell}y_j(s,z)^{b_j}\Bigg)ds,
\end{split}
\end{equation}
where $S_{\ell}$ is the set of all $\ell$-tuples of nonnegative integers $[b_1,b_2,\ldots,b_{\ell}]$ satisfying $b_1+2b_2+\cdots+\ell b_{\ell}={\ell}$ and $L=b_1+b_2+\cdots+b_{\ell}$. Here, $\partial^LF(\theta,r)$ denotes the Fr\'echet's derivative of order $L$ with respect to the variable $r$.

In \cite{jmj13,jdm14} the averaging method at any order was developed to study isolated periodic solutions of nonsmooth but continuous differential systems. Recently, the averaging method has also been extended to study isolated periodic solutions of discontinuous differential systems; see \cite{jad15,jjd17,jdc17}. The following $k$-th order averaging theorem is proved in Llibre-Novaes-Teixeira \cite{jdm14}.
\begin{theorem}\label{T1}
Assume that the following conditions hold:

(a) for each $i=0,1,\ldots,k$ and $\theta\in\mathbb{R}$, the function $F_i(\theta,\cdot)$ is of class $\mathcal{C}^{k-i}$, $\partial^{k-i}F_i$ is locally Lipschitz in the second variable, and $R(\theta,\cdot,\varepsilon)$ is a continuous function locally Lipschitz in the second variable;

(b) $f_i\equiv0$ for $i=1,2,\ldots,j-1$ and $f_j\neq0$ with $j\in\{1,2,\ldots,k\}$;

(c) for some $z^*\in D$ with $f_j(z^*)=0$, there exists a neighborhood $V\subset D$ of $z^*$ such that $f_j(z)\neq0$ for all $z\in\bar{V}\backslash\{z^*\}$, and that $d_B(f_j(z),V,0)\neq0$.

Then, for $|\varepsilon|>0$ sufficiently small, there exists a $2\pi$-periodic solution $r_{\varepsilon}(\theta)$ of \eqref{eq2.01} such that $r_{\varepsilon}(0)\rightarrow z^*$ when $\varepsilon\rightarrow 0$.
\end{theorem}

\begin{remark}\label{rem1}
The above symbol $d_B$ denotes the Browder degree; see Browder \cite{feb83} for a general definition. When $f_j$ is a $\mathcal{C}^1$ function and the derivative of $f_j$ at $z\in V$ is distinct from zero (i.e., $f_j'(z)\neq0$), then in this case, $f_j'(z^*)\neq0$ implies $d_B(f_j(z),V,0)\neq0$.
\end{remark}

Recently in \cite{dn17} the partial Bell polynomials were used to provide a relatively simple alternative formula for the recurrence \eqref{eq2.01.3}. Since the Bell polynomials are implemented in algebraic manipulators as Maple and Mathematica, this new formula can make easier the computational implementation of the averaged functions. In this paper, we will exploit this new formula in our algorithmic approach for solving the problem of evaluating the recurrence \eqref{eq2.01.3} (see Section \ref{sect3.2}). In the sequel, for $\ell$ and $m$ positive integers, we recall the Bell polynomials:
\begin{equation}\label{e3.2.1}
\begin{split}
B_{\ell,m}(x_1,\ldots,x_{\ell-m+1})=\sum_{\tilde{S}_{\ell,m}}\frac{\ell!}{b_1!b_2!\cdots b_{\ell-m+1}!}\prod_{j=1}^{\ell-m+1}\left(\frac{x_j}{j!}\right)^{b_j},\nonumber
\end{split}
\end{equation}
where $\tilde{S}_{\ell,m}$ is the set of all $(\ell-m+1)$-tuples of nonnegative integers $[b_1,b_2,\ldots,b_{\ell-m+1}]$ satisfying $b_1+2b_2+\cdots+(\ell-m+1)b_{\ell-m+1}=\ell$, and $b_1+b_2+\cdots+b_{\ell-m+1}=m$.

The following result is an equivalent formulation of the integral equation \eqref{eq2.01.3} via above Bell polynomials, its proof can be found in \cite{dn17}.

\begin{theorem}\label{T2}
For $i=1,2,\ldots,k$ the recursive equation \eqref{eq2.01.3} reads
\begin{equation}\label{eq.2.1}
\begin{split}
&y_1(\theta,z)=Y(\theta,z)\int_0^{\theta}Y(s,z)^{-1}F_1(s,r(s,z))ds,\\
&y_i(\theta,z)=Y(\theta,z)\int_0^{\theta}Y(s,z)^{-1}\Bigg(i!F_i(s,r(s,z))\\
&+\sum_{m=2}^{i}\partial^mF_0(s,r(s,z))B_{i,m}\big(y_1(s,z),\ldots,y_{i-m+1}(s,z)\big)\\
&+\sum_{\ell=1}^{i-1}\sum_{m=1}^{\ell}\frac{i!}{\ell!}\partial^mF_{i-\ell}(s,r(s,z))B_{\ell,m}\big(y_1(s,z),\ldots,y_{\ell-m+1}(s,z)\big)\Bigg)ds,
\end{split}
\end{equation}
where $Y(\theta,z)$ is the fundamental solution of the variational equation $Y'=\partial F_0(\theta,r(\theta,z))Y$
satisfying the initial condition $Y(0,z)=1$.
\end{theorem}

The general study of the exact number of simple zeros of the averaged functions \eqref{eq2.01.2} up to every order is also very difficult to be done, since the averaged functions may be too complicated, such as including square root functions, logarithmic functions, and the elliptic integrals. In the literature there is an abundance of papers dealing with zeros of the averaged functions (see, for instance, \cite{jlgs11,hjj16,bh17,bh19,bh20} and references therein). Note that one can estimate the size of bifurcated limit cycles by using the expressions of the averaged functions. In fact we know that if the averaged functions $f_j=0$ for $j=1,\ldots,k-1$ and $f_k\neq0$, and $\bar{z}\in D$ is a simple zero of $f_k$, then by Theorem \ref{T1} there is a limit cycle $r_{\varepsilon}(\theta)$ of differential system \eqref{eq2.01} such that $r_{\varepsilon}(\theta)=r(\theta,\bar{z})+\mathcal{O}(\varepsilon)$. Then, going back through the changes of variables we have for the differential system \eqref{eq2.04} the limit cycle $(x(t,\varepsilon),y(t,\varepsilon))=(r(\theta,\bar{z})\cos\theta,r(\theta,\bar{z})\sin\theta)+\mathcal{O}(\varepsilon)$.

\section{Algorithmic Averaging for the Study of Limit Cycles}\label{sect3}

The process of using the averaging method for studying limit cycles of differential systems can be divided into three steps (\cite{hy19}, Sect. 4).

{\bf STEP 1}. Write the perturbed system \eqref{eq2.04} in the normal form of averaging \eqref{eq2.01} up to $k$-th order in $\varepsilon$.

{\bf STEP 2}. (i) Compute the exact formula for the $k$-th order integral function $y_k(\theta,z)$ in \eqref{eq.2.1}. (ii) Derive the symbolic expression of the $k$-th order averaged function $f_k(z)$ by \eqref{eq2.01.2}.

{\bf STEP 3}. Determine the exact upper bound of the number of simple zeros of $f_k(z)$ for $z\in D$.

In the following subsections we will present algorithms to implement the first two steps. We use ``Maple-like'' pseudo-code, based on our Maple implementation. Using these algorithms we reduce the problem of studying the number of limit cycles of system \eqref{eq2.04} to the problem of detecting {\bf STEP 3}.

\subsection{Algorithm for transformation into normal form}\label{sect3.1}
In this subsection we will devise an efficient algorithm which can be used to transform system \eqref{eq2.04} into the form \eqref{eq2.01}.

We first describe the underlying equations before presenting the algorithm. Doing the change of polar coordinates $x=rC$, $y=rS$ with $C=\cos\theta$ and $S=\sin\theta$, then we can transform system \eqref{eq2.04} into the following form
\begin{equation}\label{eqs3.1}
\begin{split}
\frac{dr}{d\theta}&=\frac{dr/dt}{d\theta/dt}=\frac{r(x\dot{x}+y\dot{y})}{x\dot{y}-y\dot{x}}\Big|_{x=rC,y=rS}\\
&=r\frac{C\left(P(x,y)+ p(x,y,\varepsilon)\right)+S\left(Q(x,y)+ q(x,y,\varepsilon)\right)}{C\left(Q(x,y)+ q(x,y,\varepsilon)\right)-S\left(P(x,y)+ p(x,y,\varepsilon)\right)}\Big|_{x=rC,y=rS}\\
&=r\frac{\frac{CP(x,y)+SQ(x,y)}{CQ(x,y)-SP(x,y)}+\frac{Cp(x,y,\varepsilon)
+Sq(x,y,\varepsilon)}{CQ(x,y)-SP(x,y)}}{1+\frac{Cq(x,y,\varepsilon)
-Sp(x,y,\varepsilon)}{CQ(x,y)-SP(x,y)}}\Big|_{x=rC,y=rS}\\
&=F_0(\theta,r)+\varepsilon F_1(\theta,r)+\ldots+\varepsilon^k F_k(\theta,r)+\mathcal{O}(\varepsilon^{k+1}).
\end{split}
\end{equation}
The last equality is obtained by carrying the order $k+1$ Taylor series expansion of the penultimate equality, with respect to the variable $\varepsilon$, around the point $\varepsilon=0$. The first algorithm {\bf NormalForm}, presented below, is a direct implementation of the formula derivation in \eqref{eqs3.1}.
\begin{small}
\begin{algorithm}[H]
\caption{{\bf NormalForm}$(P,Q,p,q,k)$}
\hspace*{0.02in} {\bf Input:}
a perturbed system \eqref{eq2.04} with an order $k\geq0$ in \eqref{eq2.01}\\
\hspace*{0.02in} {\bf Output:}
an expression of $dr/d\theta$ up to $k$-th order in $\varepsilon$
\begin{algorithmic}[1]
\State $d1:=\mbox{normal}(\mbox{subs}(x=r\cdot C,y=r\cdot S,x\cdot(P+p)+y\cdot(Q+q))/r)$;
\State $d2:=\mbox{normal}(\mbox{subs}(x=r\cdot C,y=r\cdot S,x\cdot(Q+q)-y\cdot(P+p))/r^2)$;
\State $T:=\mbox{taylor}(d1/d2,\varepsilon=0,k+1)$;
\State $H:=\mbox{convert}\left(T,\mbox{polynom}\right)$;
\State $F_0:=\mbox{coeff}(\varepsilon\cdot H,\varepsilon)$;
\For{$i$ {\bf from} 1 {\bf to} $k$}
\State $c_i:=\mbox{coeff}(H,\varepsilon^i)$;
\State $F_{i,1}:=\mbox{prem}\left(\mbox{numer}(c_i),C^2+S^2-1,S\right)$;
\State
$F_{i,2}:=\mbox{prem}\left(\mbox{denom}(c_i),C^2+S^2-1,S\right)$;
\State
$F_i:={F_{i,1}}/{F_{i,2}}$;
\EndFor
\State $dr/d\theta:=\mbox{subs}(C=\cos\theta,S=\sin\theta,F_0+\sum_{j=1}^kF_j\varepsilon^j$);
\State \Return $dr/d\theta$;
\end{algorithmic}
\end{algorithm}
\end{small}
In line 8 the function $\mbox{prem}(a,b,x)$ is the pseudo-remainder of $a$ with respect to $b$ in the variable $x$. By the property of the pseudo-remainder we know that the degree in $S$ is at most 1 of the polynomials $F_{i,1}$ and $F_{i,2}$.

\subsection{Algorithms for computing formulae and functions of averaging}\label{sect3.2}
This subsection is devoted to provide effective algorithms to compute the formula and exact expression of the $k$-th order averaged function. According to Theorem \ref{T2}, we should take the following substeps to compute the $k$-th order averaged function of system \eqref{eq2.01}.

{\bf Substep 1}. Determine the open and bounded interval $D$, the $2\pi$-periodic solution $r(\theta,z)$ of the unperturbed system $dr/d\theta=F_0(\theta,r)$ with initial condition $r(0,z)=z\in D$, and the fundamental solution $Y(\theta,z)$ of the variational equation $Y'=\partial F_0(\theta,r(\theta,z))Y$
with initial condition $Y(0,z)=1$.

{\bf Substep 2}. Compute the exact formula for the $k$-th order integral function $y_k(\theta,z)$.

{\bf Substep 3}. Output the symbolic expression for the $k$-th order averaged function $f_k(z)$ (simplified by using the conditions for $f_1\equiv f_2\equiv\cdots\equiv f_{k-1}\equiv0$) for a given differential system \eqref{eq2.04}.

We provide each of the substep an algorithm. For the Substep 1 we first derive the $2\pi$-periodic solution $r(\theta,z)$, and then use it to further obtain the interval $D$ and the fundamental solution $Y(\theta,z)$.

\begin{algorithm}[H]
\caption{{\bf DSolutions}$(F_0)$}
\hspace*{0.02in} {\bf Input:}
the unperturbed term $F_0$ in \eqref{eq2.01}\\
\hspace*{0.02in} {\bf Output:}
$r(\theta,z)$, a set of inequalities ($SIs$) with respect to $z$, and $Y(\theta,z)$
\begin{algorithmic}[1]
\State $\mbox{de1}:=\mbox{diff}(r(\theta),\theta)=\mbox{subs}(r=r(\theta),F_0)$;
\State $\mbox{ans1}:={\bf dsolve}(\{\mbox{de1},r(0)=z\},r(\theta))$;
\State $r(\theta,z):=\mbox{op}(2,\mbox{ans1})$;
\State $\mbox{minvalue}:={\bf minimize}(r(\theta,z),\theta=0..2\pi)$;
\State $m:=\mbox{nops}([\mbox{op}(\mbox{minvalue})])$;
\State $SIs:=\{\}$;
\For{$i$ {\bf from} 1 {\bf to} $m$}
\State $SIs:=SIs ~{\bf union}~ \{\mbox{op}(i,\mbox{minvalue})>0\}$;
\EndFor
\State $\mbox{de2}:=\mbox{diff}(Y(\theta),\theta)=\mbox{subs}(r=r(\theta,z),\mbox{diff}(F_0,r))\cdot Y(\theta)$;
\State $\mbox{ans2}:={\bf dsolve}(\{\mbox{de2},Y(0)=1\},Y(\theta))$;
\State $Y(\theta,z):=\mbox{op}(2,ans2)$;
\State \Return [$r(\theta,z)$, $SIs$, $Y(\theta,z)$];
\end{algorithmic}
\end{algorithm}

\begin{remark}\label{rem3.2}
The output results of $r(\theta,z)$ and $Y(\theta,z)$ can be reduced by using the identity $\sin^2\theta+\cos^2\theta=1$ so that the degree of what are left in $\sin\theta$ is at most 1. We use the routine {\bf dsolve} built-in Maple for solving an ordinary differential equation. We remark that the unperturbed term $F_0(\theta,r)$ is usually a rational trigonometric function in $r$, $\sin\theta$ and $\cos\theta$. As far as we know, we do not have a systematic approach to the solution of the differential equation $dr/d\theta=F_0(\theta,r)$ in the general case. In Section \ref{sect4} we will consider certain classes of differential systems with uniform isochronous centers to illustrate the effectiveness of our algorithm. It is important to emphasize that the interval $D$ can be determined by using the output set $SIs$. Since the original system may contain some parameters, the resulting set $SIs$ could be parametric. In order to derive the interval $D$ in this case, we will construct an equivalent solution set $\overline{SIs}$ of $SIs$ that contains only the rational polynomial inequalities, and then use the {\bf SemiAlgebraic} command in Maple to compute the solutions. Below we provide a concrete example to show the feasibility this algorithm, one may check the results in \cite{hjj16}. More experiments can be found in Section \ref{sect4}.
\end{remark}

\begin{example}\label{exa.1}
Consider the following quintic polynomial differential system
\begin{equation}\label{e.q3.2.1}
\begin{split}
\dot{x}=-y+x^2y(x^2+y^2),\quad \dot{y}=x+xy^2(x^2+y^2).
\end{split}
\end{equation}
Applying our algorithm {\bf NormalForm} for $p=q=k=0$, we have $dr/d\theta=r^5\cos\theta\sin\theta$. Then applying the algorithm {\bf DSolutions} we obtain a list [$r(\theta,z)$, $SIs$, $Y(\theta,z)$], where
\begin{equation}\label{e.q3.2.2}
\begin{split}
r(\theta,z)&=\frac{z}{(2z^4(\cos^2\theta-1)+1)^{1/4}},\\
SIs&=\left\{0<z,0<\frac{z}{(-2z^4+1)^{1/4}}\right\},\\
Y(\theta,z)&=\frac{1}{(2z^4(\cos^2\theta-1)+1)^{5/4}}.\nonumber
\end{split}
\end{equation}
To obtain the interval $D$ in this case, we construct an equivalent solution set $\overline{SIs}$ of $SIs$ that contains only the rational polynomials: $\overline{SIs}:=\{0<z,0<\frac{z}{-2z^4+1}\}$. Then using the Maple command {\bf SolveTools[SemiAlgebraic]}, we compute the solution of the set $\overline{SIs}$, and obtain that $D=\{0<z<2^{-1/4}\}$.
\end{example}

We want to say that the expressions of the returned results on $r(\theta,z)$ and $Y(\theta,z)$ may be complicated, such as including square root functions, and exponential functions. Below we give a simple example to show this, one may find the related results in \cite{syz15}.
\begin{example}\label{exa.2}
Consider the following polynomial differential system
\begin{equation}\label{e.q3.2.1}
\begin{split}
\dot{x}=-y(3x^2+y^2),\quad \dot{y}=x(x^2-y^2).
\end{split}
\end{equation}
The normal form $dr/d\theta=-2r\cos\theta\sin\theta$ can be obtained by the algorithm {\bf NormalForm}, and the algorithm {\bf DSolutions} returns a list $\big[ze^{\cos^2\theta-1},\big\{0<z,0<ze^{-1}\big\},
e^{\cos^2\theta-1}\big]$.
\end{example}

For the Substep 2, we present our algorithm {\bf AveragingFormula}. This algorithm can be used to compute the exact formula of the $k$-th order integral function $y_k(\theta,z)$. Correctness of it follows from Theorem \ref{T2}.

\begin{algorithm}[H]
\caption{{\bf AveragingFormula}$(k)$}
\hspace*{0.02in} {\bf Input:}
an order $k\geq1$ of the normal form \eqref{eq2.01}\\
\hspace*{0.02in} {\bf Output:}
the integral function $y_k(\theta,z)$
\begin{algorithmic}[1]
\State $T_1:=0$; $T_2:=0$;

\For{$m$ {\bf from} 2 {\bf to} $k$}

\State $T_1:=T_1+\mbox{Diff}(F_0(s,r(s,z)),r\$m)\cdot\mbox{{\bf IncompleteBellB}}(k,m,y_1(s,z),\ldots,y_{k-m+1}(s,z))$;

\EndFor

\For{$\ell$ {\bf from} 1 {\bf to} $k-1$}
\For{$m$ {\bf from} 1 {\bf to} $\ell$}

\State $T_2:=T_2+\frac{k!}{\ell!}\cdot\mbox{Diff}
(F_{k-\ell}(s,r(s,z)),r\$m)\cdot\mbox{{\bf IncompleteBellB}}(\ell,m,y_1(s,z),\ldots,y_{\ell-m+1}(s,z))$;
\EndFor
\EndFor
\State
\begin{small}
$y_k(\theta,z):=Y(\theta,z)\cdot\mbox{Int}\left(Y^{-1}(s,z)\cdot\left(k!\cdot F_k(s,r(s,z))+T_1+T_2\right),s=0..\theta\right)$;
\end{small}
\State \Return $y_k(\theta,z)$;
\end{algorithmic}
\end{algorithm}
We deduce explicitly the formulae of $y_k$'s up to $k=5$ in Appendix \ref{A}. In fact our algorithm can compute arbitrarily high order formulae of $y_k$'s. In Section \ref{sect4}, we will study several differential systems to show the feasibility of our algorithm.

In the last subsection, we provide an algorithm {\bf NormalForm} to transform system \eqref{eq2.04} into the form $dr/d\theta$. The algorithm {\bf DSolutions} admits one to obtain the fundamental solutions
$r(\theta,z)$, $Y(\theta,z)$ and the interval $D$ (Substep 1). The algorithm {\bf AveragedFunction}, presented below, is based on the algorithms {\bf NormalForm}, {\bf DSolutions} and Theorem \ref{T2}, which provides a straightforward calculation method to derive the exact expression of the $k$-th order averaged function for a given differential system in the form \eqref{eq2.04} (Substep 3).
\begin{small}
\begin{algorithm}[H]
\caption{{\bf AveragedFunction}$(dr/d\theta,r(\theta,z),Y(\theta,z),k)$}
\hspace*{0.02in} {\bf Input:}
a normal formal of averaging \eqref{eqs3.1} with an order $k\geq1$ and the fundamental solutions $r(\theta,z)$, $Y(\theta,z)$\\
\hspace*{0.02in} {\bf Output:}
a list of expressions of the averaged functions
\begin{algorithmic}[1]

\State $F_0:=\mbox{coeff}(\varepsilon\cdot(dr/d\theta),\varepsilon)$;

\For{$j$ {\bf from} 1 {\bf to} $k$}

\State $F_j:=\mbox{coeff}(dr/d\theta,\varepsilon^j)$;

\State $A_j:={\bf AFormula}(j)$;

\State $H_j:=\mbox{normal}\left(\frac{1}{Y(\theta,z)}
\cdot\text{expand}(\mbox{subs}(r=r(\theta,z),\text{value}(A_j)))\right)$;

\State $H_{j,1}:=\mbox{collect}(\mbox{expand}(\mbox{numer}(H_j)),\{\cos\theta,\sin\theta\},\mbox{distributed})$;

\State $H_{j,2}:=\text{denom}(H_j)$;

\For{$h$ {\bf from} 1 {\bf to} $\text{nops}(H_{j,1})$}

\State $g_{j,h}:=\mbox{int}\left(\frac{\mbox{op}(h,H_{j,1})}{H_{j,2}},\theta=0..\theta,\mbox{AllSolutions}\right)$;

\State $s_{j,h}:=\mbox{int}\left(\frac{\mbox{op}(h,H_{j,1})}{H_{j,2}},\theta=0..2\pi\right)$;

\EndFor

\State $y_j:=Y(\theta,z)\cdot\mbox{sum}(g_{j,t},t=1..\text{nops}(H_{j,1}))$;

\State $f_j:=\frac{1}{j!}\cdot\mbox{sum}(s_{j,t},t=1..\text{nops}(H_{j,1}))$;

\EndFor

\State \Return [$y_k$, $f_k$];
\end{algorithmic}
\end{algorithm}
\end{small}
In line 4, the routine {\bf AFormula} is a subalgorithm we use for the generation of the expression in the parenthesis of equation \eqref{eq.2.1} without dependence on $(s,z)$. The detailed information of this subalgorithm is as follows.\\
{\bf Subalgorithm: AFormula}\\
INPUT: An averaging order $k\geq1$;\\
OUTPUT: The expression in the parenthesis of equation \eqref{eq.2.1} without dependence on $(s,z)$.\\
STEP 0. $U=0$; $V=0$;\\
STEP 1. {\bf For} $m$ {\bf from} 2 {\bf to} $k$ do

\qquad\quad$U:=U+\text{Diff}(F_0,r\$m)\cdot{\bf IncompleteBellB}(k,m,$

\qquad\quad$\text{seq}(y_i,i=1..k-m+1))$; {\bf end do};\\
STEP 2. {\bf For} $\ell$ {\bf from} 1 {\bf to} $k-1$ do

\qquad\quad{\bf for} $m$ {\bf from} 1 {\bf to} $\ell$ do

\qquad\quad$V:=V+\frac{k!}{\ell!}\cdot\text{Diff}(F_{k-\ell},r\$m)\cdot{\bf IncompleteBellB}(\ell,m,$

\qquad\quad$\text{seq}(y_i,i=1..\ell-m+1))$; {\bf end do}; {\bf end do};\\
STEP 3. Output $k!F_k+U+V$.

\begin{remark}\label{rem3.3}
In order to obtain an exact and simplified expression of the averaged function, one should make some assumptions (e.g., the interval $D$ on $z$, and possible conditions on the parameters that may appear in the original differential systems) before preforming the algorithm {\bf AveragedFunction}. For more details see our experiments in Section \ref{sect4}. We also remark that, throughout the computation, an assumption on $\theta$ (i.e., $\theta\in(2\pi-\epsilon,2\pi+\epsilon)$ with $\epsilon$ a small number) was made to identify a valid branch of the possible returned piecewise functions (in line 9), since the integral functions $y_i(\theta,z)$ for $i=1,\ldots,k$ evaluate at the point $\theta=2\pi$ in \eqref{eq2.01.2}.

\end{remark}

We implemented all the algorithms presented in this section in Maple. In the next section, we will apply our general algorithmic approach to analyze the bifurcation of limit cycles for several concrete differential systems.

\section{Implementation and Experiments}\label{sect4}
In this section we demonstrate our algorithmic tests using several examples. We present the bifurcation of limit cycles for a class of generalized Kukles polynomial differential systems as an illustration of our approach explained in previous sections. In addition, we study the number of limit cycles that bifurcate from some periodic solutions surrounding the isochronous centers for certain differential systems by the first and second order averaging method. The obtained results of our experiments extend the existing relevant results and show the feasibility of our approach.

\subsection{A class of generalized Kukles differential systems}\label{sect4.1}
In this subsection we consider a very particular case of the 16th Hilbert problem; we study the number of limit cycles of the generalized Kukles polynomial differential system
\begin{equation}\label{ex.q4.1}
\begin{split}
\dot{x}=-y,\quad \dot{y}=x+Q(x,y),
\end{split}
\end{equation}
where $Q(x,y)$ is a polynomial with real coefficients of degree $n$. This system was introduced by Kukles in \cite{isk44}, examining the conditions under which the origin of the system
\begin{equation}\label{ex.q4.2}
\begin{split}
\dot{x}&=-y,\\ \dot{y}&=x+a_1x^2+a_2xy+a_3y^2+a_4x^3+a_5x^2y+a_6xy^2+a_7y^3\nonumber
\end{split}
\end{equation}
is a center. For long time, it had been thought that the conditions given by Kukles were necessary and sufficient conditions, but some new cases have been found, see \cite{jw90,cjng90}.

Here we are interested in studying the maximum number of limit cycles that bifurcate from the periodic orbits of the linear center $\dot{x}=-y$, $\dot{y}=x$, perturbed inside the following class of generalized Kukles polynomial differential systems
\begin{equation}\label{ex.q4.3}
\begin{split}
\dot{x}&=-y+\sum_{k\geq1}\varepsilon^kl_m^k(x),\quad \\ \dot{y}&=x-\sum_{k\geq1}\varepsilon^k\left(f_{n_1}^k(x)+g_{n_2}^k(x)y+h_{n_3}^k(x)y^2+d_0^ky^3\right),
\end{split}
\end{equation}
where for every $k$ the polynomials $l_m^k(x)$, $f_{n_1}^k(x)$, $g_{n_2}^k(x)$, and $h_{n_3}^k(x)$ have degree $m$, $n_1$, $n_2$, and $n_3$ respectively, $d_0^k\neq0$ is a real number and $\varepsilon$ is a small parameter. This question has been studied in \cite{naa19} for $k=1,2$, and the authors obtained the following result.
\begin{theorem}\label{T4.1}
Assume that for $k=1,2$ the polynomials $l_m^k(x)$, $f_{n_1}^k(x)$, $g_{n_2}^k(x)$, and $h_{n_3}^k(x)$ have degree $m$, $n_1$, $n_2$, and $n_3$ respectively, with $m$, $n_1$, $n_2$, $n_3\geq1$, and $d_0^k\neq0$ is a real number. Then for $\varepsilon$ sufficiently small the maximum number of limit cycles of the Kukles polynomial system \eqref{ex.q4.3} bifurcating from the periodic orbits of the linear center $\dot{x}=-y$, $\dot{y}=x$,
\begin{enumerate}
  \item is $\max\left\{\left[\frac{m-1}{2}\right],\left[\frac{n_2}{2}\right],1\right\}$ by using the first order averaging method;
  \item is $\max\big\{\left[\frac{n_1}{2}\right]+\left[\frac{n_2-1}{2}\right],\left[\frac{n_1}{2}\right]+\left[\frac{m}{2}\right]-1,
      \left[\frac{n_1+1}{2}\right],\left[\frac{n_3+3}{2}\right],\\
      \left[\frac{n_3}{2}\right]+\left[\frac{m}{2}\right],
      \left[\frac{n_2+1}{2}\right]+\left[\frac{n_3}{2}\right],\left[\frac{n_2}{2}\right],
      \left[\frac{m-1}{2}\right],\left[\frac{n_1-1}{2}\right]+\mu,\\ \left[\frac{n_3+1}{2}\right]+\mu,1\big\}$ by using the second order averaging method, where $\mu=\min\left\{\left[\frac{m-1}{2}\right],\left[\frac{n_2}{2}\right]\right\}$.
\end{enumerate}

\end{theorem}
Here, [$\cdot$] denotes the integer part function. Remark that, many researchers have discussed the bifurcation of limit cycles for generalized Kukles polynomial differential system in the form \eqref{ex.q4.1}. We refer the readers to \cite{jlam11,arc18} for some interesting results on this subject. The next result extends Theorem \ref{T4.1} to arbitrary order of averaging.
\begin{lemma}\label{L4.2}
Let $\max\{m,n_1,n_2+1,n_3+2\}=N\geq3$, then the Kukles polynomial system \eqref{ex.q4.3} for $\varepsilon$ sufficiently small has no more than $\left[{k(N-1)}/{2}\right]$ limit cycles bifurcating from the periodic orbits of the linear center $\dot{x}=-y$, $\dot{y}=x$, using the averaging method up to order $k$.
\end{lemma}

\begin{proof}
This result follows directly from Theorem 6 in \cite{jmj13}.
\end{proof}

In what follows, using our algorithms we will do some experimental results by fixing some values of the degrees in system \eqref{ex.q4.3}. Note that the maximum numbers of limit cycles in Theorem \ref{T4.1} and Lemma \ref{L4.2} may not be reached. The following corollary shows that these maximum numbers can be reached for some orders of averaging.

\begin{corollary}\label{cork.1}
(i) When $m=3$, $n_1=3$, $n_2=2$, and $n_3=1$, the maximum number of limit cycles of the Kukles polynomial system \eqref{ex.q4.3} bifurcating from the periodic orbits of the linear center $\dot{x}=-y$, $\dot{y}=x$, using the fifth order averaging method is five and it is reached.

(ii) When $m=5$, $n_1=1$, $n_2=2$, and $n_3=1$, the maximum number of limit cycles of the Kukles polynomial system \eqref{ex.q4.3} bifurcating from the periodic orbits of the linear center $\dot{x}=-y$, $\dot{y}=x$, using the fourth order averaging method is five and it is reached.
\end{corollary}

The detailed proof of the first statement of Corollary \ref{cork.1} can be found in Appendix \ref{A-B}. Since the calculations and arguments of the second part are quite similar to those used in the first one, we omit the proof of statement (ii) in Corollary \ref{cork.1}. More concretely, we provide in Table \ref{tb01} the maximum number of limit cycles for system \eqref{ex.q4.3} in each case of Corollary \ref{cork.1} up to the $k$-th order averaging method for $k=1,\ldots,5$.

\begin{table}[h]
\caption{Number of limit cycles of system \eqref{ex.q4.3} in Corollary \ref{cork.1}}\label{tb01}
\begin{center}
\begin{tabular}{ccc}
  \hline
  Averaging order & Statement (i) & Statement (ii)\\ \hline
   1 & 1 & 2\\ \hline
   2 & 2 & 2\\ \hline
   3 & 3 & 4\\ \hline
   4 & 4 & 5\\ \hline
   5 & 5 & -\\ \hline
\end{tabular}
\end{center}
\end{table}
The number of limit cycles in statement (i) can be reached for each order of averaging. That is to say, the bound given in Lemma \ref{L4.2} is sharp for the case in statement (i). However, for the statement (ii), the bound given in Lemma \ref{L4.2} is only sharp for the first order of averaging. We note also that for each statement in Corollary \ref{cork.1}, the bound provided in Theorem \ref{T4.1} can be reached up to the second order.
\begin{remark}\label{remcor.1}
The calculation of the high order averaged function $f_k$ involves heavy computations with complicated expressions. It may not work effectively when one of the degrees ($m$ and $n_i$, $i=1,2,3$) is big. It turns out that we can greatly improve the speed by updating the obtained $dr/d\theta$ using the conditions on the parameters of $f_1\equiv f_2\equiv\cdots\equiv f_{k-1}=0$.
\end{remark}

\subsection{Limit cycles for certain differential systems with uniform isochronous centers}\label{sect4.2}
Recall that a center $\bar{x}$ of system \eqref{eq2.02} is an \textit{isochronous center} if it has a neighborhood such that in this neighborhood all the periodic orbits have the same period. An isochronous center is \textit{uniform} if in polar coordinates $x=r\cos\theta$, $y=r\sin\theta$, it can be written as $\dot{r}=G(\theta,r)$, $\dot{\theta}=\eta$, $\eta\in\mathbb{R}\backslash\{0\}$, see Conti \cite{rc94} for more details. The next result on the uniform isochronous center (UIC) is well-known, a proof of it can be found in \cite{jijl15}.
\begin{proposition}\label{prop1}
Assume that system \eqref{eq2.02} has a center at the origin $\bar{x}$. Then $\bar{x}$ is a UIC if and only if by doing a linear change of variables and a rescaling of time the system can be written as
\begin{equation}\label{ex2.1}
\begin{split}
\dot{x}=-y+xf(x,y),\quad \dot{y}=x+yf(x,y),
\end{split}
\end{equation}
where $f$ is a polynomial in $x$ and $y$ of degree $n-1$, and $f(0,0)=0$.
\end{proposition}
In what follows, we recall some important results on the UICs of planar cubic and quartic differential systems. The following result due to Collins \cite{cbc97} in 1997, also obtained by Devlin, Lloyd and Pearson \cite{jnlj98} in 1998, and by Gasull, Prohens and Torregrosa \cite{arj05} in 2005 characterizes the UICs of cubic polynomial systems.
\begin{theorem}\label{thex.1}
A planar cubic differential system has a UIC at the origin if and only if it can be written as system \eqref{ex2.1} with $f(x,y)=a_1x+a_2y+a_3x^2+a_4xy-a_3y^2$ satisfying that $a_1^2a_3-a_2^2a_3+a_1a_2a_4=0$. Moreover, this planar cubic differential system can be reduced to either one of the following two forms:
\begin{eqnarray}\label{ex2.2}
&&\dot{x}=-y+x^2y,\quad\dot{y}=x+xy^2, \label{ex2.2.1}\\
&&\dot{x}=-y+x^2+Ax^2y,\quad\dot{y}=x+xy+Axy^2, \label{ex2.2.2}
\end{eqnarray}
where $A\in\mathbb{R}$.
\end{theorem}
Systems \eqref{ex2.2.1} and \eqref{ex2.2.2} are known as \textit{Collins First Form} and \textit{Collins Second Form}, respectively. See (\cite{jj15}, Thm. 9) for more details of the global phase portraits of the Collins forms.

In order to save space, we put the remaining results in Appendix \ref{B}.

\section{Conclusion}\label{sect5}
We have presented a systematical approach to analyze how many limit cycles of differential system \eqref{eq2.04} can bifurcate from the periodic orbits of an unperturbed one via the averaging method. We designed four algorithms to analyze the averaging method and shown that the general study of the number of limit cycles of system \eqref{eq2.04} can be reduced to the problem of estimating the number of simple zeros of the obtained averaged functions with the aid of these algorithms.

Our algorithms admit a generalization to the case of studying the bifurcation of limit cycles for discontinuous differential systems. It would be interesting to employ our approach to analyze the bifurcation of limit cycles for differential systems in many different fields, which are of high interest in nature sciences and engineering. It will be beneficial to generalize our current approach to the case of higher dimension differential systems by using the general form of the averaging method. We leave this as the future research problems.

In addition, we noticed the phenomenon of tremendous growth of expressions in intermediate calculations while we done experiments for the linear center $\dot{x}=-y$, $\dot{y}=x$ by using the high order of averaging. For the nonlinear polynomial differential centers, the evaluation of the high order averaged functions is highly nontrivial; the main difficulty exists in the technical and cumbersome computations of some complicated integral equations. How to simplify and optimize the steps of the computations of the averaged functions is also a question that remains for further investigation.

\section*{Acknowledgments}
Huang's work is partially supported by China Scholarship Council under Grant No.:~201806020128. The author is grateful to Professor Chee Yap and Professor Dongming Wang for their profound concern and encouragement.

\bibliographystyle{plain}
\bibliography{ref}

\appendix

\section{Fifth Order Averaging Formulae}\label{A}
The explicit formulae of the functions $y_k(\theta,z)$ for $k=1,2,\ldots,5$.
\begin{equation}
\begin{split}
y_1(\theta,z)&=Y\left(\theta,z\right)\int_{0}^{\theta}{\frac{F_{{1}} \left(s,r\left(s,z\right)\right)}{Y\left(s,z\right)}}{ds},\\
y_2(\theta,z)&=Y\left(\theta,z\right)\int_{0}^{\theta}\frac{1}{Y\left(s,z
\right)}\Bigg(2F_{{2}}
\left(s,r\left(s,z\right)\right)+{\frac{\partial^{2}}
{\partial{r}^{2}}}F_{{0}}\left(s,r\left(s,z\right)\right)
y_{{1}}\left(s,z\right)^{2}\\
&+2{\frac{\partial }{\partial r}}F_{{1}}\left(s,r\left(s,z\right)\right)y_{{1}} \left(s,z\right)\Bigg){ds},\\
y_3(\theta,z)&=Y\left(\theta,z\right)\int_{0}^{\theta}\frac{1}{Y\left(s,z
\right)}\Bigg(6F_{{3}}\left(s,r\left(s,z\right)\right)+3{\frac{\partial^
{2}}{\partial{r}^{2}}}F_{{0}}\left(s,r\left(s,z\right)\right)y_{{1}} \left(s,z\right)y_{{2}}\left(s,z\right)\\
&+{\frac{\partial^{3}}{\partial{r}^{3}}}F_{{0}}\left(s,r
\left(s,z\right)\right)y_{{1}}\left(s,z\right)^{3}+6{\frac{\partial }{\partial r}}F_{{2}}\left(s,r\left(s,z\right)\right)y_{{1}}\left(s,z\right)\\
&+3{\frac{\partial}{\partial r}}F_{{1}}\left(s,r\left(s,z\right)\right) y_{{2}}\left(s,z\right)+3{\frac{\partial^{2}}{\partial{r}^{2}}}F_{{1}}\left(s,r\left(s,z
\right)\right)y_{{1}}\left(s,z\right)^{2}\Bigg){ds},\\
y_4(\theta,z)&=Y\left(\theta,z\right)\int _{0}^{\theta}\frac{1}{Y\left(s,z
\right)}\Bigg(24F_{{4}}\left(s,r\left(s,z\right)\right)\\
&+{\frac{\partial ^{2}}{\partial{r}^{2}}}F_{{0}}\left(s,r\left(s,z\right)\right)
\left(4y_{{1}}\left(s,z\right)y_{{3}}\left(s,z\right)+3y_{{2}}
\left(s,z\right)^{2}\right)\\
&+6{\frac{\partial^{3}}{\partial{r}^{3}}}F_{{0}}\left(s,
r\left(s,z\right)\right)y_{{1}}\left(s,z\right)^{2}y_{{2}}\left(s,z
\right)+{\frac{\partial^{4}}{\partial{r}^{4}}}F_{{0}}\left(s,r\left(s,z
\right)\right)y_{{1}}\left(s,z\right)^{4}\\
&+24{\frac{\partial}{\partial
r}}F_{{3}}\left(s,r\left(s,z\right)\right)y_{{1}}\left(s,z\right)
+12{\frac{\partial}{\partial r}}F_{{2}}\left(s,r\left(s,z\right)\right)  y_{{2}}\left(s,z\right)\\
&+4{\frac{\partial}{\partial r}}F_{{1
}}\left(s,r\left(s,z\right)\right)y_{{3}}\left(s,z\right)+12{\frac{\partial^{2}}{\partial{r}^{2}}}F_{{2}} \left(s,r\left(s,z\right)\right)y_{{1}}\left(s,z\right)^{2}\\
&+4{\frac{\partial^{3}}
{\partial{r}^{3}}}F_{{1}}\left(s,r\left(s,z\right)\right)y_{{1}}\left(s,z\right)^{3}
+12{\frac{\partial^{2}}{\partial{r}^{2}}}F_{{1}}
\left(s,r\left(s,z\right)\right)y_{{1}}\left(s,z\right)y_{{2}}\left(s,z
\right)\Bigg){ds},\\
y_5(\theta,z)&=Y\left(\theta,z\right)\int _{0}^{\theta}\frac{1}{Y\left(s,z
\right)}\Bigg(120F_{{5}}\left(s,r\left(s,z\right)\right)
+5{\frac{\partial^{2}}{\partial{r}^{2}}}F_{{0}}\left(s,r\left(s,z\right)\right)y_{{1}}\left(s,z\right)y_{{4}}\left(s,z\right)\\
&+10{\frac{\partial^{2}}{\partial{r}^{2}}}F_{{0}}\left(s,r\left(s,z\right)\right)y_{{2}}\left(s,z\right)y_{{3}}\left(s,z\right)
+10{\frac{\partial^{3}}{\partial{r}^{3}}}F_{{0}}\left(s,r\left(s,z\right) \right)y_{{1}}\left(s,z\right)^{2}y_{{3}}\left(s,z\right)\\
&+15{\frac{\partial^{3}}{\partial{r}^{3}}}F_{{0}}\left(s,r\left(s,z\right) \right)y_{{1}}\left(s,z\right)y_{{2}}\left(s,z\right)^{2}
+10{\frac{\partial^{4}}{\partial{r}
^{4}}}F_{{0}}\left(s,r\left(s,z\right)\right)y_{{1}}\left(s,z\right) ^{3}y_{{2}}\left(s,z\right)\\
&+{\frac{\partial^{5}}{\partial {r}^{5}}}F_{{0}}\left(s,r\left(s,z\right)\right)y_{{1}}\left(s,z\right)^{5}
+120{\frac{\partial}{\partial r}}F_{{4}}\left(s,r\left(s,z\right) \right)y_{{1}}\left(s,z\right)\\
&+60{\frac{\partial}{\partial r}}F_{{3}} \left(s,r\left(s,z\right)\right)y_{{2}}\left(s,z\right)
+60{\frac{\partial^{2}}{\partial{r}^{2}}}F_{{3}}\left(s,r\left(s,z
\right)\right)y_{{1}}\left(s,z\right)^{2}\\
&+20{\frac{\partial}{\partial r}}F_{{2}}\left(s,r\left(s,z\right)\right)y_{{3}}\left(s,z\right)
+60{\frac{\partial^{2}}{\partial{r}^{2}}}F_{{2}}\left(s,r\left(s,z
\right)\right)y_{{1}}\left(s,z\right)y_{{2}}\left(s,z\right)\\
&+20{\frac
{\partial^{3}}{\partial{r}^{3}}}F_{{2}}\left(s,r\left(s,z\right)\right)  y_{{1}}\left(s,z\right)^{3}
+5{\frac{\partial^{2}}{\partial{r}^{2}}}F_{{1}}\left(s,r\left(s,z\right) \right)\left(4y_{{1}}\left(s,z\right)y_{{3
}}\left(s,z\right)+3y_{{2}}\left(s,z\right)^{2}\right)\\
&+30{\frac{\partial^{3}}{\partial{r}^{3}}}F_{{1}}\left(s,r\left(s,z \right)\right)y_{{1}}\left(s,z\right)^{2}y_{{2}}\left(s,z\right)\\
&+5{\frac{\partial}{\partial r}}F_{{1}}\left(s,r\left(s,z\right)\right)y_{{4}}\left(s,z\right)+5{\frac{\partial^{4}}{\partial{r}^{4}}}F_{{1}}\left(s,r\left(s,z\right) \right)y_{{1}}\left(s,z\right)^{4}\Bigg){ds}.
\end{split}
\end{equation}

\section{Proof of Corollary \ref{cork.1}}\label{A-B}
Let
\begin{equation}\label{A-B.1}
\begin{split}
l_3^k(x)&=\sum_{i=0}^3e_{k,i}x^i,\quad f_3^k(x)=\sum_{i=0}^3a_{k,i}x^i,\\
g_2^k(x)&=\sum_{i=0}^2b_{k,i}x^i,\quad h_1^k(x)=\sum_{i=0}^1c_{k,i}x^i.
\end{split}
\end{equation}
We consider the following perturbed system
\begin{equation}\label{A-B.2}
\begin{split}
\dot{x}&=-y+\sum_{k=1}^5\varepsilon^kl_3^k(x),\quad \\ \dot{y}&=x-\sum_{k=1}^5\varepsilon^k\left(f_{3}^k(x)+g_{2}^k(x)y+h_{1}^k(x)y^2+d_0^ky^3\right).
\end{split}
\end{equation}
Now we study the number of limit cycles of system \eqref{A-B.2}. Applying our algorithm {\bf NormalForm} by taking $k=5$ we obtain
\begin{equation}\label{A-B.3}
\begin{split}
\frac{d r}{d\theta}=\sum_{i=1}^5\varepsilon^i F_i(\theta,r)+\mathcal{O}(\varepsilon^6).
\end{split}
\end{equation}
Here we give only the expression of $F_1(\theta,r)$, the explicit expressions of $F_i(\theta,r)$ for $i=2,\ldots,5$ are quite large so we omit them.
\begin{equation}\label{A-B.4}
\begin{split}
F_1(\theta,r)&=S\Big((-{C}^{3}a_{{1,3}}+{C}^{3}c_{{1,1}}-Cc_{{1,1}})r^3
+(-{C}^{2}a_{{1,2}}+{C}^{2}c_{{1,0}}-c_{{1,0}})r^2-Ca_{1,1}r-a_{1,0}\Big)\\
&+\Big(({C}^{4}b_{{1,2}}-{C}^{4}d_{{1,0}}+{C}^{4}e_{{1,3}}-{C}^{2}b_{{1,2}}+2{C}^{2}d_{{1,0}}-d_{{1,0}})r^3\\
&+({C}^{3}b_{{1,1}}+{C}^{3}e_{{1,2}}-Cb_{{1,1}})r^2
+({C}^{2}b_{{1,0}}+{C}^{2}e_{{1,1}}-b_{{1,0}})r+Ce_{1,0}\Big)\nonumber
\end{split}
\end{equation}
with $C=\cos\theta$ and $S=\sin\theta$. For the case $F_0=0$, the algorithm {\bf DSolutions} returns a list [$z,\{0<z\},1$]. We obtain the first function by using the algorithm {\bf AveragedFunction}
\begin{equation}\label{A-B.5}
\begin{split}
f_1(z)&=-\frac{\pi z}{4}\Big((b_{{1,2}}+3d_{{1,0}}-3e_{{1,3}})z^2+4(b_{{1,0}}-e_{{1,1}})\Big).\nonumber
\end{split}
\end{equation}
Therefore $f_1(z)$ can have at most one positive simple real root. From Theorem \ref{T1} it follows that the first order averaging provides the existence of at most one limit cycle of system \eqref{A-B.2} and this number can be reached since the coefficients in $f_1(z)$ are independent constants.

From now on, for each $k=2,\ldots,5$, we will perform the calculation of the averaged function $f_k$ under the hypothesis $f_j\equiv0$ for $j=1,\ldots,k-1$.

Doing $e_{{1,3}}=d_{{1,0}}+b_{{1,2}}/3$, $e_{{1,1}}=b_{{1,0}}$, and computing $f_2$ we obtain
\begin{equation}\label{A-B.6}
\begin{split}
f_2(z)&=-\frac{\pi z}{24}\Big(A_{2,4}z^4+A_{2,2}z^2+A_{2,0}\Big),\nonumber
\end{split}
\end{equation}
where
\begin{equation}\label{A-B.7}
\begin{split}
A_{2,4}&=-9a_{{1,3}}d_{{1,0}}+2b_{{1,2}}c_{{1,1}}+3c_{{1,1}}d_{{1,0}},\\
A_{2,2}&=-18a_{{1,1}}d_{{1,0}}+6a_{{1,2}}b_{{1,1}}-12a_{{1,2}}e_{{1,2}}+12
b_{{1,0}}c_{{1,1}}\\
&\quad+6b_{{1,1}}c_{{1,0}}+6b_{{2,2}}+18d_{{2,0}}-
18e_{{2,3}},\\
A_{2,0}&=24a_{{1,0}}b_{{1,1}}-48a_{{1,0}}e_{{1,2}}+48c_{{1,0}}e_{{1,0}}+
24b_{{2,0}}-24e_{{2,1}}.\nonumber
\end{split}
\end{equation}
It is obvious that $f_2(z)$ can have at most two positive simple real roots. Therefore, the second order averaging provides the existence of at most two limit cycles of system \eqref{A-B.2} and this number can be reached.

Letting $e_{2,1}=A_{2,0}/24+e_{2,1}$, $e_{2,3}=A_{2,2}/18+e_{2,3}$, $a_{1,3}=A_{2,4}/9d_{1,0}+a_{1,3}$, and computing $f_3$ we have
\begin{equation}\label{A-B.8}
\begin{split}
f_3(z)&=\frac{\pi z}{864d_{1,0}}\Big(A_{3,6}z^6+A_{3,4}z^4+A_{3,2}z^2+A_{3,0}\Big),\nonumber
\end{split}
\end{equation}
where $A_{3,6}=3d_{1,0}(5b_{1,2}+12d_{1,0})(3b_{1,2}d_{1,0}-c_{1,1}^2)$, the expressions of $A_{3,2i}$ for $i=0,1,2$ are quite long, so we do not provide them here. It is not hard to check that $f_3(z)$ has at most three positive simple real roots. Then the third order averaging provides the existence of at most three limit cycles of system \eqref{A-B.2} and this number can be reached.

To consider the fourth order averaging theorem we have two cases from the expression of $A_{3,6}$.

{\bf CASE 1}: [$5b_{1,2}+12d_{1,0}=0$], and {\bf CASE 2}: [$3b_{1,2}d_{1,0}-c_{1,1}^2=0$].

In what follows we only show the proof under the {\bf CASE 1}, because the maximum number of limit cycles in Lemma \ref{L4.2} under this case can be reached up to $k=5$ for system \eqref{A-B.2}.

In order to let $f_3\equiv0$ we take $e_{3,1}=-A_{3,0}/864d_{1,0}+e_{3,1}$, $e_{3,3}=-A_{3,2}/648d_{1,0}+e_{3,3}$, $a_{2,3}=-A_{3,4}/324d_{1,0}^2+a_{2,3}$, $b_{1,2}=-12d_{1,0}/5$. Computing $f_4$ we obtain
\begin{equation}\label{A-B.9}
\begin{split}
f_4(z)&=-\frac{\pi z}{216000d_{1,0}}\Big(A_{4,8}z^8+A_{4,6}z^6+A_{4,4}z^4+A_{4,2}z^2+A_{4,0}\Big),\nonumber
\end{split}
\end{equation}
where $A_{4,8}=27c_{1,1}d_{1,0}^2(5c_{1,1}^2+36d_{1,0}^2)$, here again we omit the expressions of $A_{4,2i}$ for $i=0,1,2,3$, because they are too long. Hence $f_4(z)$ has at most four positive simple roots. Then the fourth order averaging provides the existence of at most four limit cycles of system \eqref{A-B.2} and this number can be reached.

Note that $d_{1,0}\neq0$, so in order to let $f_4\equiv0$, we take $c_{1,1}=0$ and the following equalities
\begin{equation}\label{A-B.10}
\begin{split}
e_{4,1}&=A_{4,0}/216000d_{1,0}+e_{4,1},\quad
e_{4,3}=A_{4,2}/162000d_{1,0}+e_{4,3},\\
a_{3,3}&=A_{4,4}/81000d_{1,0}^2+a_{3,3},\quad
b_{2,2}=-A_{4,6}/750d_{1,0}(5c_{1,1}^2+36d_{1,0}^2)+b_{2,2}.\nonumber
\end{split}
\end{equation}

Computing $f_5$ we obtain
\begin{equation}\label{A-B.11}
\begin{split}
f_5(z)=-\frac{\pi z}{19440000d_{1,0}}\Big(A_{5,10}z^{10}+A_{5,8}z^8+A_{5,6}z^6+A_{5,4}z^4+A_{5,2}z^2+A_{5,0}\Big),\nonumber
\end{split}
\end{equation}
where $A_{5,10}=40824d_{1,0}^6$, and
\begin{equation}\label{A-B.12}
\begin{split}
A_{5,8}&=-243d_{1,0}^4\Big(10375{a_{{1,2}}}^{2}+16870a_{{1,2}}c_{{1,0}}-1080b_{{1,0}}d_{{1,0}}
+2451{b_{{1,1}}}^{2}
\\&\quad+18456b_{{1,1}}e_{{1,2}}+2715{c_{{1,0}}}^{2}-12291{e_{{1,2}}}^{2}-360c_{{2,1}}\Big).\nonumber
\end{split}
\end{equation}

We do not explicitly provide the expressions of $A_{5,2i}$ for $i=0,1,2,3$, because they are very long. Therefore $f_5(z)$ can have at most five positive real roots. Then the fifth order averaging provides the existence of at most five limit cycles of system \eqref{A-B.2} and this number can be reached.

\section{Limit cycles for certain differential systems with uniform isochronous centers}\label{B}
This appendix is an overflow from Subsection \ref{sect4.2}. The following characterization of planar quartic polynomial differential systems with an isolated UIC at the origin is provided by Chavarriga, Garc\'ia and Gin\'e \cite{jaj01}, in 2001.
\begin{theorem}\label{thex.2}
A planar quartic differential system has a UIC at the origin if and only if it can be written as
\begin{equation}\label{ex2.3}
\begin{split}
\dot{x}&=-y+x\left(ax+bxy+cx^3+dxy^2\right),\\
\dot{y}&=x+y\left(ax+bxy+cx^3+dxy^2\right),
\end{split}
\end{equation}
where $a,b,c,d\in\mathbb{R}$.
\end{theorem}
A classification of the global phase portraits of the quartic differential systems of the form \eqref{ex2.3} is provided in \cite{jijl15}.

In this subsection, we study the limit cycles that bifurcate from some periodic orbits surrounding the origin of the Collins First Form and a subclass of system \eqref{ex2.3} (we take $a=b=0$), respectively.

\subsection{Bifurcation of limit cycles of Collins First Form}\label{sect4.2.1}
Consider the following perturbation of the Collins First Form:
\begin{equation}\label{ex4.2.1}
\begin{split}
\dot{x}&=-y+x^2y+\varepsilon\sum_{i+j=0}^3a_{i,j}x^iy^j,\\
\dot{y}&=x+xy^2+\varepsilon\sum_{i+j=0}^3b_{i,j}x^iy^j.
\end{split}
\end{equation}
We study the number of limit cycles of system \eqref{ex4.2.1} by using the first order averaging method. Applying our algorithm {\bf NormalForm} by taking $k=1$ we obtain
\begin{equation}\label{ex4.2.2}
\begin{split}
\frac{dr}{d\theta}=F_0(\theta,r)+\varepsilon F_1(\theta,r)+\mathcal{O}(\varepsilon^2),
\end{split}
\end{equation}
where $F_0(\theta,r)=r^3\cos\theta\sin\theta$, and
\begin{equation}\label{ex4.2.3}
\begin{split}
F_1(&\theta,r)=S\Big({r}^{5}\left(a_{{0,3}}-a_{{2,1}}+b_{{1,2}}-b_{{
3,0}}\right){C}^{5}
+{r}^{4}\left(-a_{{1,1}}+b_{{0,2}}-b_{{2,0}}\right){C}^{4}\\
&+\big({r}^{5}(-2a_{{0,3}}+a_{{2,1}}-b_{{1,2}})
+{r}^{3}(-a_{{0,1}}-a_{{0,3}}+a_{{2,1}
}-b_{{1,0}}-b_{{1,2}}+b_{{3,0}})\big){C}^{3}\\
&+\left({r}^{4}(a_{{1,1}}-b_{{0,2}})+{r}^{2}(a_{{1,1}}-b_{{0
,0}}-b_{{0,2}}+b_{{2,0}})\right){C}^{2}+\big({r}^{5}a_{{0,3}}+{r}^{3}(a_{{0,1}}+a_{{0,3}}\\
&+b_{{1,2}})+r(a_{{0,1}}+b_{{1,0}})\big)C+{r}^{2}b_{{0,2}}+b_{{0,0}}
\Big)+{r}^{5}\left(a_{{1,2}}-a_{{3,0}}-b_{{0,3}}+b_{{
2,1}}\right){C}^{6}\\
&+{r}^{4}\left(a_{{0,2}}-a_{{2,0}}+b_{{1,1}}\right){C}^{5}
+\big({r}^{5}(-2a_{{1,2}}+a_{{3,0}}+2b_{{0,3}}-b_{{2,1}})\\
&+{r}^{3}(-a_{{1,0}}-a_{{1,
2}}+a_{{3,0}}+b_{{0,1}}+b_{{0,3}}-b_{{2,1}})\big){C}^{4}+\big({r}^{4}(-2a_{{0,2}}+a_{{2,0}}-b_{{1,1}})\\
&+{r}^{2}(-a_{{0,0}}
-a_{{0,2}}+a_{{2,0}}-b
_{{1,1}})\big){C}^{3}+\big({r}^{5}(a_{{1,2}}-b_{{0,3}})\\
&+{r}^{3}(a_{{1,0}}+a_{{1,2}}-b_{{0,1}}-2b_{{0,3}}+b_{{2,1}})
+r(a_{{1,0}}-b_{{0,1}})\big){C}^{2}\\
&+\big({r}^{4}a_{{0,2}}+{r}^{2}(a_{{0,0}}+a_{{0,2}}+b_{{1,1}})+a_{{0,0}}
\big)C+{r}^{3}b_{{0,3}}+rb_{{0,1}}\nonumber
\end{split}
\end{equation}
with $S=\sin\theta$, $C=\cos\theta$. Using the algorithm {\bf DSolutions} we obtain a list [$r(\theta,z)$,$SIs$,$Y(\theta,z)$], where
\begin{small}
\begin{equation}\label{ex4.2.4}
\begin{split}
r(\theta,z)=\frac{z}{\sqrt{z^2(\cos^2\theta-1)+1}},\quad SIs&=\left\{0<z,0<\frac{z}{\sqrt{-z^2+1}}\right\},\quad
Y(\theta,z)=\frac{1}{(z^2(\cos^2\theta-1)+1)^{3/2}}.\nonumber
\end{split}
\end{equation}
\end{small}
To obtain the interval $D$ in this case, we construct an equivalent solution set $\overline{SIs}:=\{0<z,0<\frac{z}{-z^2+1}\}$.
Computing the solution of the set $\overline{SIs}$, we have $D=\{0<z<1\}$. Now assuming $z\in D$ and applying the algorithm {\bf AveragedFunction} for the system \eqref{ex4.2.1}, we obtain
\begin{equation}\label{ex4.2.5}
\begin{split}
f_1(z)&=-\frac{\pi}{z}\Big(\big((b_{0,1}-b_{0,3}-b_{2,1})z^4+(-a_{1,0}
-3a_{3,0}-b_{0,1}-b_{0,3}+a_{1,2}+3b_{2,1})z^2\\
&\quad+2a_{3,0}+2b_{0,3}-2a_{1,2}-2b_{2,1}\big)+2\sqrt{1-z^2}\\
&\quad\times\big((a_{3,0}-b_{2,1})z^2-a_{3,0}-b_{0,3}+a_{1,2}+b_{2,1}\big)\Big).\nonumber
\end{split}
\end{equation}
By taking the transformation $z=\sqrt{1-s^2}$, $0<s<1$, we then have
\begin{equation}\label{ex4.2.6}
\begin{split}
zf_1(z)&=\pi(1-s)\Big((b_{{0,1}}-b_{{0,3}}-b_{{2,1}}){s}^{3}+(b_{{0,1}}-2a_{{
3,0}}-b_{{0,3}}+b_{{2,1}}){s}^{2}+(a_{{1,0}}-a_{{1,2}}\\
&\quad+a_{{3,0}}+2b_{{0,3}})s+a_{{1,0}}+a_{{1,2}}+a_{{3,0}}\Big).\nonumber
\end{split}
\end{equation}
It is not hard to check that $f_1(z)$ can have 3 simple zeros in the interval $(0,1)$. Thus system \eqref{ex4.2.1} can have up to 3 limit cycles by using the first order averaging method from Theorem \ref{T1}. Therefore, using our algorithmic approach we verified the first result in (\cite{hjj16}, Thm. 1.1). In the following, we provide a concrete example to demonstrate that there exist some systems in the form \eqref{ex4.2.1} having exactly three limit cycles. Consider a family of systems
\begin{equation}\label{A-C.1}
\begin{split}
\dot{x}&=-y+x^2y+\varepsilon\Big((-26-b_{0,3}-b_{2,1})x+(\frac{61}{2}+b_{2,1})x^3
+(-\frac{11}{2}+b_{0,3})xy^2\Big),\\
\dot{y}&=x+xy^2+\varepsilon\Big((30+b_{2,1}+b_{0,3})y+b_{2,1}x^2y+b_{0,3}y^3\Big).
\end{split}
\end{equation}
Applying our algorithmic approach to system \eqref{A-C.1}, we obtain the following first order averaged function
\begin{equation}\label{A-C.2}
\begin{split}
f_1(z)\overset{z=\sqrt{1-s^2}}{\Longrightarrow}\pi\frac{\sqrt{1-s}}{\sqrt{1+s}}\big((3s-1)(2s-1)(5s-1)\big),
\end{split}
\end{equation}
where $z$ and $s$ are defined as before. Apparently, $f_1(z)$ has exactly three positive zeros, denoted by
\[z_1=\frac{\sqrt{3}}{2},\quad z_2=\frac{2\sqrt{2}}{3},\quad z_3=\frac{2\sqrt{6}}{5},\]
corresponding to $s_1=1/2$, $s_2=1/3$, and $s_3=1/5$, respectively, in $z\in(0,1)$. Then it follows from Theorem \ref{T1} that system \eqref{A-C.1} has exactly three limit cycles by the first order averaging method. We explicitly provide the expressions of these three limit cycles as follows:
\begin{equation}\label{A-C.3}
\begin{split}
r_1\left(\theta,\frac{\sqrt{3}}{2}\right)=\frac{\sqrt{3}}{\sqrt{3\cos^2\theta+1}},\quad
r_2\left(\theta,\frac{2\sqrt{2}}{3}\right)=\frac{2\sqrt{2}}{\sqrt{8\cos^2\theta+1}},\quad
r_3\left(\theta,\frac{2\sqrt{6}}{5}\right)=\frac{2\sqrt{6}}{\sqrt{24\cos^2\theta+1}}.\nonumber
\end{split}
\end{equation}

\begin{remark}
(i) Usually for the nonlinear differential systems, the higher the averaging order is, the more complex are the computational operations to calculate the averaged functions. Even for the first order averaging, if one consider a perturbation of a given differential system inside the class of polynomial differential systems of the same degree (see \cite{jijl2015,hjj16,hbnw19}), the calculations of the averaged function require powerful computational resources. In next subsection, we will try to analyze a class of quartic systems by the second order averaging method. (ii) To our best knowledge, the bifurcation of medium amplitude limit cycles for the Collins Second Form has not been studied by the averaging method. Here we attempt to do this, but our algorithm {\bf DSolutions} can not return the desired results. In fact, for the Collins Second Form, one can obtain the unperturbed normal form: $dr/d\theta=Ar^3\cos\theta\sin\theta+r^2\cos\theta$ by the algorithm {\bf NormalForm}. However, determining the explicitly fundamental solutions $r(\theta,z)$ and $Y(\theta,z)$ for this parametric differential system $dr/d\theta$ seems to be very hard.
\end{remark}

\subsection{Bifurcation of limit cycles of system \eqref{ex2.3}$|_{a=b=0}$}\label{sect4.2.2}
Consider system \eqref{ex2.3} with $a=b=0$:
\begin{equation}\label{ex4.3.0}
\begin{split}
\dot{x}=-y+x(cx^3+dxy^2),\quad\dot{y}=x+y(cx^3+dxy^2).
\end{split}
\end{equation}
First, note that we may use a spatial scaling $x\rightarrow x/\sqrt[3]{d}$, $y\rightarrow y/\sqrt[3]{d}$ ($d\neq0$) to system \eqref{ex4.3.0} to obtain
\begin{equation}\label{ex4.3.1}
\begin{split}
\dot{x}=-y+x(\alpha x^3+xy^2),\quad\dot{y}=x+y(\alpha x^3+xy^2),
\end{split}
\end{equation}
where $\alpha=c/d$. We now consider the following class of perturbation of system \eqref{ex4.3.1}:
\begin{equation}\label{ex4.3.2}
\begin{split}
\dot{x}&=-y+x(\alpha x^3+xy^2)+\sum_{k=1}^2\varepsilon^k\Big(\lambda_kx+\sum_{i+j=4}a_{k,i,j}x^iy^j\Big),\\
\dot{y}&=x+y(\alpha x^3+xy^2)+\sum_{k=1}^2\varepsilon^k\Big(\lambda_ky+\sum_{i+j=4}b_{k,i,j}x^iy^j\Big),
\end{split}
\end{equation}
where $\lambda_k$, $a_{k,i,j}$'s and $a_{k,i,j}$'s are real constants. Here we study the number of limit cycles of system \eqref{ex4.3.2} by the second order averaging method. We remark that special case of system \eqref{ex4.3.2} ($\alpha=1$, $\lambda_1=0$) has been studied by the first order averaging method (see \cite{lf14}), In fact, the system there can be written as the form $\eqref{ex4.3.1}|_{\alpha=1}$ by doing a linear change of variables.

Now applying our algorithm {\bf NormalForm} by taking $k=2$ we obtain
\begin{equation}\label{ex4.3.3}
\begin{split}
\frac{dr}{d\theta}=F_0(\theta,r)+\varepsilon F_1(\theta,r)+\varepsilon^2 F_2(\theta,r)+\mathcal{O}(\varepsilon^3),
\end{split}
\end{equation}
where $F_0(\theta,r)=r^4\cos\theta\left((\alpha-1)\cos^2\theta+1\right)$, and
\begin{equation}\label{ex4.3.4}
\begin{split}
F_1(\theta,r)&=S\Big({r}^{7}(\alpha-1)(a_{{1,0,4}}-a_{{1,2,2}}+a_{{1,4,0}}+b_{{1,1,3}}
-b_{{1,3,1}})C^7-{r}^{7}\big(\alpha(2a_{{1,0,4}}-a_{{1,2,2}}+b_{
{1,1,3}})\\
&-3a_{{1,0,4}}+2a_{{1,2,2}}-a_{{1,4,0}}-2b_{{1,1,3}}+b_{{1,3,1}}\big)C^5-{r}^{4}\big(a_{{1,1,3}}-a_{{1,3,1}}-b_{{1,0,4}}
+b_{{1,2,2}}\\
&-b_{{1,4,0}}\big)C^4+{r}^{7}\left(\alpha a_{{1,0,4}}-3a_{{1,0,4}}+a_{{1,2,2}}-b_{{1,1,3}}\right)C^3
+{r}^{4}\left(a_{{1,1,3}}-2b_{{1,0,4}}+b_{{1,2,2}}\right)C^2\\
&+{r}^{7}a_{{1,0,4}}C+r^4b_{1,0,4}\Big)
+{r}^{7}(\alpha-1)(a_{{1,1,3}}-a_{{1,3,1}}-b_{{1,0,4}}+b_{{1,2,2}}-b_{{1,4,0}})C^8\\
&-{r}^{7}(\alpha(2a_{{1,1,3}}-a_{{1,3,1}}-2b_{{1,0,4}}+b_{{1,2,2}})-3a_{{1,1,3}}
+2a_{{1,3,1}}+3b_{{1,0,4}}-2b_{{1,2,2}}+b_{{1,4,0}})C^6\\
&+{r}^{4}\big(a_{{1,0,4}}-a_{{1,2,2}}+a_{{1,4,0}}+b_{{1,1,3}}-b_{{1,3,1}}\big)C^5+{r}^{7}\big(\alpha(a_{{1,1,3}}-b_{{1,0,4}})\\
&-3a_{{1,1,3}}+a_{{1,3,1}}+3b_{{1,0,4}}-b_{{1,2,2}}\big)C^4
-{r}^{4}\left(2a_{{1,0,4}}-a_{{1,2,2}}+2b_{{1,1,3}}-b_{{1,3,1}}\right)C^3\\
&+{r}^{7}\left(a_{{1,1,3}}-b_{{1,0,4}}\right)C^2
+{r}^{4}\left(a_{{1,0,4}}+b_{{1,1,3}}\right)C+r\lambda_1\nonumber
\end{split}
\end{equation}
with $S=\sin\theta$, $C=\cos\theta$, the expression of $F_2(\theta,r)$ is quite long so we omit it here. Using the algorithm {\bf DSolutions} we obtain a list [$r(\theta,z),SIs,Y(\theta,z)$], where
\begin{small}
\begin{equation}\label{ex4.3.5}
\begin{split}
r(\theta,z)&=\frac{z}{\left[1-(\alpha-1)z^3\cos^2\theta\sin\theta-(2\alpha+1)z^3\sin\theta\right]^{1/3}},\\ SIs&=\Big\{0<z,0<\frac{z}{(1-(2\alpha+1)z^3)^{1/3}}, 0<\frac{z}{(1+(2\alpha+1)z^3)^{1/3}}\Big\},\\
Y(\theta,z)&=\frac{1}{\left[1-(\alpha-1)z^3\cos^2\theta\sin\theta-(2\alpha+1)z^3\sin\theta\right]^{4/3}}.\nonumber
\end{split}
\end{equation}
\end{small}
To obtain the interval $D$ in this case, we construct an equivalent solution set $\overline{SIs}:=\{0<z,0<\frac{z}{1-(2\alpha+1)z^3},0<\frac{z}{1+(2\alpha+1)z^3}\}$.
By using the Maple command {\bf SolveTools[SemiAlgebraic]} to $\overline{SIs}$, we obtain the interval $D$ as follows:
\begin{equation}\label{ex4.3.6}
D=\left\{\begin{split}
\Big\{0<z,z<-\frac{1}{(2\alpha+1)^{1/3}}\Big\}&,\quad \mbox{if}~\alpha<-1/2,\\
\{0<z\}&,\quad \mbox{if}~\alpha=-1/2,\\
\Big\{0<z,z<\frac{1}{(2\alpha+1)^{1/3}}\Big\}&,\quad \mbox{if}~\alpha>-1/2.\\
\end{split}\right.\nonumber
\end{equation}
Now assuming the case $\alpha>-1/2$ and applying the algorithm {\bf AveragedFunction} for the system \eqref{ex4.3.2} when $k=1$, we find that
Maple may not evaluate the following kind of parametric integrals:
\begin{equation}\label{ex4.3.7}
\begin{split}
I_{i,j}=\int_0^{2\pi}\left(\frac{\cos^i\theta\sin^j\theta}{1-(\alpha-1)z^3\cos^2\theta\sin\theta-(2\alpha+1)z^3\sin\theta}\right)d\theta,\quad i,j\in\mathbb{N}.\nonumber
\end{split}
\end{equation}
In view of this, from now on we restrict to the case $\alpha=1$. Computing the first averaged function by the algorithm {\bf AveragedFunction} and taking the transformation $z=\left[\frac{1-s^2}{3(1+s^2)}\right]^{1/3}$ for $0<s<1$, we have
\begin{equation}\label{ex4.3.7}
\begin{split}
f_1(z)&\overset{z=\left[\frac{1-s^2}{3(1+s^2)}\right]^{1/3}}{\Longrightarrow}\bar{f}_1(s)\\
&=\frac{\pi3^{2/3}(1-s)^{1/3}}{108(1+s)^{11/3}(1+s^2)^{4/3}}\big(N_1s^6+N_2s^5
+N_3s^4+N_4s^3+N_3s^2+N_2s+N_1\big),
\end{split}
\end{equation}
where
\begin{equation}\label{ex4.3.8}
\begin{split}
N_1&=3a_{{1,1,3}}+a_{{1,3,1}}-3b_{{1,0,4}}-b_{{1,2,2}}-3b_{{1,4,0}}+
72\lambda_{{1}},\\
N_2&=-4a_{{1,1,3}}+4b_{{1,0,4}}+4a_{{1,3,1}}-4b_{{1,2,2}}-12b_{{1
,4,0}}+288\lambda_{{1}},\\
N_3&=5a_{{1,1,3}}-9a_{{1,3,1}}-5b_{{1,0,4}}+9b_{{1,2,2}}-5b_{{1,4
,0}}+504\lambda_{{1}},\\
N_4&=-8a_{{1,1,3}}+8a_{{1,3,1}}+8b_{{1,0,4}}-8b_{{1,2,2}}+40b_{{1
,4,0}}+576\lambda_{{1}}.\nonumber
\end{split}
\end{equation}
As a result of the symmetry of coefficients of the function $\bar{f}_1(s)$ with respect to $s$, we know that if $s_0\neq0$ is one root of $\bar{f}_1(s_0)$, so is $1/s_0$. Hence, the fact that $\bar{f}_1(s)$ has at most three zeros in $s\in(0,1)$ implies that there exist at most three zeros for $f(z)$ in $z\in(0,1/\sqrt[3]{3})$. Moreover, the number three can be reached since the constants $N_1$, $N_2$, $N_3$ and $N_4$ are independent. It follows from Theorem \ref{T1} that the first order averaging provides the existence of at most three limit cycles of system \eqref{ex4.3.2}, and this number can be reached. Hence, with the nonzero constant $\lambda_1$ the perturbed system \eqref{ex4.3.2} can produce one more limit cycle than the case without it.

In order to consider the second order averaging, we must use the conditions of $f_1(z)=0$. The following lemma follows from equation \eqref{ex4.3.7}.
\begin{lemma}\label{lemsec4.3}
The first averaged function $f_1(z)\equiv0$ if and only if $a_{1,1,3}=b_{1,0,4}$, $a_{1,3,1}=b_{1,2,2}$, and $b_{1,4,0}=\lambda_1=0$.
\end{lemma}
Now we consider the second order averaging of system \eqref{ex4.3.2} with $\alpha=1$. First, we update the obtained $dr/d\theta$ in \eqref{ex4.3.3} by Lemma \ref{lemsec4.3}. We put the resulting expression of $y_1(\theta,z)$ as follows.
\begin{equation}\label{B.1}
\begin{split}
y_1(\theta,z)&=\frac{1}{14580z^{11}(3z^3\sin\theta-1)^{4/3}}\Big(m_1\cdot\ln(1-3z^3\sin\theta)\\
&\quad+m_2\cdot\cos^2\theta\sin\theta+m_3\cdot\cos^4\theta\sin\theta+m_4\cdot\cos^4\theta\\
&\quad+m_5\cdot\cos^3\theta+m_6\cdot\cos^2\theta-14580b_{1,0,4}z^{15}\cos\theta\\
&\quad+m_7\cdot\sin\theta+m_8\Big),
\end{split}
\end{equation}
where
\begin{equation}\label{B.2}
\begin{split}
m_1&=1620\left(-a_{{1,4,0}}+b_{{1,3,1}}\right){z}^{12}+180
\big(-a_{{1,2,2}}+2a_{{1,4,0}}\\
&\quad+b_{{1,1,3}}-2b_{{1,3,1}}\big){z}^{6}+20(a_{{1,2,2}}-a_{{1,0,4}}-a_{{1,4,0}}-b_{{1,1,3}}+b_{{1,3,1}}),\\
m_2&=-36z^9\Big(9\big(12a_{{1,0,4}}-2a_{{1,2,2}}-8a_{{1,4,0}}+17b_{{1,1,3}}\\
&\quad-7b_{{1,3,1}}\big)z^6+5(a_{{1,2,2}}-a_{{1,0,4}}-a_{{1,4,0}}
-b_{{1,1,3}}+b_{{1,3,1}})\Big),\\
m_3&=1944z^{15}(b_{{1,1,3}}-a_{{1,2,2}}-b_{{1,3,1}}+a_{{1,4,0}}+a_{{1,0,4}}),\\
m_4&=-405z^{12}(b_{{1,1,3}}-a_{{1,2,2}}-b_{{1,3,1}}+a_{{1,4,0}}+a_{{1,0,4}}),\\
m_5&=4860z^{15}(b_{{1,0,4}}-b_{{1,2,2}}),\\
m_6&=90z^6\Big(9\big(a_{{1,0,4}}-a_{{1,4,0}}+b_{{1,3,1}}\big)z^6+b_{{1,1,3}}-a_{{1,2,2}}
-b_{{1,3,1}}+a_{{1,4,0}}+a_{{1,0,4}}\Big),\\
m_7&=12z^3\Big(27(6a_{{1,0,4}}+4a_{{1,2,2}}+16a_{{1,4,0}}+11b_{{1,1,3}}
+14b_{{1,3,1}})z^{12}-15(a_{{1,0,4}}+2a_{{1,2,2}}\\
&\quad-5a_{{1,4,0}}-2b_{{1,1,3}}+5b_{{1,3,1}})z^6+5(a_{{1,2,2}}-a_{{1,0,4}}-a_{{1,4,0}}-b_{{1,1,3}}+b_{{1,3,1}})\Big),\\
m_8&=45z^6\Big(108\big(b_{{1,2,2}}+2b_{{1,0,4}}\big)z^9+9\big(-3b_{{1,3,1}}-a_{{1,2,2}}
+b_{{1,1,3}}-a_{{1,0,4}}+3a_{{1,4,0}}\big)z^6\\
&\quad+2(b_{{1,3,1}}-b_{{1,1,3}}+a_{{1,2,2}}-a_{{1,4,0}}-a_{{1,0,4}})\Big).\nonumber
\end{split}
\end{equation}

Applying the algorithm {\bf AveragedFunction} to the updated normal form $dr/d\theta$ when $k=2$, we find that Maple was consuming too much of the CPU during a calculation, and can not able to allocate enough memory. This is mainly because the expressions involved in the analysis are huge, and sometimes Maple can not evaluate the integrals for certain complicated functions. For instance, in our case here, Maple can not evaluate the integrals in $z\in(0,1/\sqrt[3]{3})$ of the following form:
\begin{equation}\label{ex4.3.9}
\begin{split}
J_{i}=\int_0^{2\pi}\left(\frac{\ln\left(1-3z^2\sin\theta\right)\cos^{2i}\theta}{9z^6(\cos^2\theta-1)+6z^3\sin\theta-1}\right)d\theta,\quad i\in\mathbb{N}.\nonumber
\end{split}
\end{equation}
To simplify our computation, we let the term $\ln\left(1-3z^2\sin\theta\right)$ in $y_1(\theta,z)$ be identically zero (i.e., $m_1\equiv0$ in \eqref{B.1}). The following corollary follows from this fact.
\begin{corollary}\label{cor.sec.4}
The term $\ln\left(1-3z^2\sin\theta\right)$ in $y_1(\theta,z)$ vanishes if and only if $a_{1,2,2}=b_{1,1,3}$, $a_{1,4,0}=b_{1,3,1}$, and $a_{1,0,4}=0$.
\end{corollary}
Applying Lemma \ref{lemsec4.3} and Corollary \ref{cor.sec.4} to system \eqref{ex4.3.2}, we obtain the following second order averaged function by the algorithm {\bf AveragedFunction}.
\begin{equation}\label{ex4.3.10}
\begin{split}
f_2(z)&=-\frac{\pi}{2916z^{11}\sqrt{1-9z^6}}\Big((h_1z^{12}+h_2z^6+8h_3)\sqrt{1-9z^6}\\
&\quad-8(1-9z^6)\big(81b_{2,4,0}z^{12}+h_4z^6+h_3\big)\Big),\nonumber
\end{split}
\end{equation}
where
\begin{equation}\label{ex4.3.11}
\begin{split}
h_1&=81a_{{2,1,3}}+243a_{{2,3,1}}-81b_{{2,0,4}}-243b_{{2,2,2}}
+1215b_{{2,4,0}}-5832\lambda_2,\\
h_2&=36a_{{2,1,3}}-108a_{{2,3,1}}-36b_{{2,0,4}}+108b_{{2,2,2}}-180b_{{2,4,0}},\\
h_3&=-a_{{2,1,3}}+a_{{2,3,1}}+b_{{2,0,4}}-b_{{2,2,2}}+b_{{2,4,0}},\\
h_4&=-9a_{{2,3,1}}+9b_{{2,2,2}}-18b_{{2,4,0}}.\nonumber
\end{split}
\end{equation}
After making the transformation as before, $f_2(z)$ becomes
\begin{equation}\label{ex4.3.12}
\begin{split}
f_2(z)&=\frac{\pi3^{2/3}(1-s)^{1/3}}{108(1+s)^{11/3}(1+s^2)^{4/3}}\big(H_1s^6+H_2s^5
+H_3s^4+H_4s^3+H_3s^2+H_2s+H_1\big),
\end{split}
\end{equation}
where
\begin{equation}\label{ex4.3.13}
\begin{split}
H_1&=3a_{{2,1,3}}+a_{{2,3,1}}-3b_{{2,0,4}}-b_{{2,2,2}}-3b_{{2,4,0}}+
72\lambda_{{2}},\\
H_2&=-4a_{{2,1,3}}+4b_{{2,0,4}}+4a_{{2,3,1}}-4b_{{2,2,2}}-12b_{{2
,4,0}}+288\lambda_{{2}},\\
H_3&=5a_{{2,1,3}}-9a_{{2,3,1}}-5b_{{2,0,4}}+9b_{{2,2,2}}-5b_{{2,4
,0}}+504\lambda_{{2}},\\
H_4&=-8a_{{2,1,3}}+8a_{{2,3,1}}+8b_{{2,0,4}}-8b_{{2,2,2}}+40b_{{2
,4,0}}+576\lambda_{{2}}.\nonumber
\end{split}
\end{equation}
It follows from Theorem \ref{T1} and equation \eqref{ex4.3.12} that the averaging method up to second order provides the existence of at most three limit cycles of system \eqref{ex4.3.2}, and this number can be reached by using similar arguments to the function $f_1(z)$.

We summarize our results as follows based on the above analysis.
\begin{theorem}\label{thse4.1}
For sufficiently small parameter $|\varepsilon|>0$, the perturbed system \eqref{ex4.3.2} with $\alpha=1$ has at most three limit cycles bifurcating from periodic orbits of the unperturbed one by the first order averaging method, and by the second order averaging method under the condition: $a_{1,2,2}=b_{1,1,3}$, $a_{1,4,0}=b_{1,3,1}$, and $a_{1,0,4}=0$. In each case this number can be reached.
\end{theorem}

\end{document}